\documentclass{article}

\usepackage{PRIMEarxiv}

\usepackage[utf8]{inputenc} 
\usepackage[T1]{fontenc}    
\usepackage{hyperref}       
\usepackage{url}            
\usepackage{booktabs}       
\usepackage{amsfonts}       
\usepackage{nicefrac}       
\usepackage{microtype}      
\usepackage{lipsum}
\usepackage{fancyhdr}       
\usepackage{graphicx}       
\usepackage{caption}
\usepackage{subcaption}

\title{PODPose: Integrating Proper Orthogonal Decomposition and EITPose
}

\author{
  Jessie Sheflin\\
  Biomedical Engineering \\
  Northwestern University \\
  Evanston, IL\\
  \texttt{\{JessieSheflin2025\}@u.northwestern.edu} \\
}

\begin{document}
\maketitle

\begin{abstract}
This work examines two ways of using proper orthogonal decomposition (POD) to enhance the prior work of EITPose, a device which uses electrical impedance tomography (EIT) to detect posture by way of a band of electrodes on the forearm. First, an electrode placement algorithm is described, which employs the sensitivity volume method and a POD basis to choose the combination of electrode locations that spans the POD basis most effectively. Next, a data placement algorithm is introduced, which uses a POD basis to account for electrodes that are providing poor data. Analysis is conducted on these two algorithms using the same techniques as the original EITPose paper, and it is shown that the electrode placement has little effect, but the data projection algorithm is very accurate when synthesizing data. The data projection algorithm represents a novel technique for adapting EIT devices live to poor electrodes, and can be applied to future implementations of the sensing technique.
\end{abstract}


\section{Introduction}
Electrical impedance tomography (EIT) is a promising method for imaging and sensing in human-computer interaction, as it is continuous, low-bandwidth, and much more private than a sensing method like cameras. 
EIT uses an array of electrodes to measure the conductance of a region of interest (ROI) by taking a number of four-point resistance measurements, which can then be used in machine learning or typical inverse methods to discern relevant information from the conductivity of the ROI  \cite{harikumar2013electrical}. 
This study concerns EITPose, a device that uses EIT to sense a user's gestures based on impedance measurements made by a band of electrodes on their forearm \cite{Kyu_Mao_Zhu_Goel_Ahuja_2024}. By bringing contemporary algorithms used in EIT such as the sensitivity volume method and proper orthogonal decomposition (POD) to EITPose, this paper seeks to improve the form factor and fidelity of the device \cite{Onsager_Wang_Costakis_Aygen_Lang_Lee_Grayson_2024}\cite{Lipponen_Seppanen_Kaipio_2013}. POD is applied to EITPose in two main ways: an algorithm to improve electrode placement, and an algorithm to compensate for electrodes with poor or no signal. First, the sensitivity volume method is used to evaluate the fidelity of different possible electrode placements using a basis consisting of the modes of the POD of data collected in the original EITPose study. Next, a data projection method is presented that uses the POD of the training data to account for missing impedance measurements. These two interventions were evaluated using the same methods used in the EITPose paper. Based on this evaluation, the electrode placement intervention has little to no improvement over evenly spaced electrodes but the data projection method is shown to accurately compensate for missing electrodes.

\section{Background}
\label{sec:blgd}

\subsection{The EIT Problem}
In EIT, electrodes are placed on the boundary of the region of interest, and four-point measurements are taken, which means that two of the electrodes are selected as positive and negative current terminals, and two more electrodes are selected as positive and negative voltage terminals \cite{vdp}. Each of $D$ total measurements uses a unique permutation of four electrodes, and the voltage values are combined into the data vector $\mathbf{d}.$ 

Finite element methods have become a popular and useful way to solve EIT problems, most notably EIDORS, the software used in this analysis \cite{Adler_Lionheart_2006}. In these programs, the ROI is modeled as $M$ small triangles, where the conductivities of each triangle compose $\mathbf{m},$ the model vector. It's possible to describe the relationship between $\mathbf{d}$ and $\mathbf{m}$ as:
\begin{equation}
        \mathbf{Jm} = \mathbf{d},
    \end{equation}
where $\mathbf{J}$ is the Jacobian, a large matrix that has $M$ columns and $D$ rows \cite{Gómez-Laberge_2008}. Each element $J_{ik}$ in the Jacobian describes the impact the change in conductivity of the $k^\mathrm{th}$ mesh piece $m_k$ has on the $i^\mathrm{th}$ measurement $d_i$. 

The maximum possible number of independent measurements $D_0$ is related to the number of contacts $C$ \cite{BHBrown_1987} through the equation 
\begin{equation}
    D_0 = \frac{C(C-3)}{2}.
\end{equation}
Two measurements where the voltage pair for one is the current pair of the other both make the same voltage measurement thanks to Onsager reciprocity; these two measurements are referred to as "Onsager pairs". The EITPose device uses $2D_0$ measurements, as it includes both members of $D_0$ Onsager pairs.

\subsection{EITPose}
This works builds heavily on EITPose, which takes impedance measurements using a band of electrodes on the forearm alongside hand position detection using Mediapipe, and trains a random forest model to reconstruct hand position from impedance data \cite{Kyu_Mao_Zhu_Goel_Ahuja_2024} \cite{Lugaresi_Tang_Nash_McClanahan_Uboweja_Hays_Zhang_Chang_Yong_Lee_etal._2019}. The same armband from the original study was used in this paper, and much of the same code was utilized. Additionally, the evaluation of this paper uses the same code as the within-session evaluation from EITPose. EITPose employs the "skip" measurement set, which uses each possible combination of two adjacent current electrodes and two adjacent voltage electrodes. This is not the optimal measurement protocol in terms of signal-to-noise ratio, but it is acceptable (~85\% of best signal-to-noise) \cite{onsager2024sensitivity}.

\subsection{Proper Orthogonal Decomposition}
Proper orthogonal decomposition is a method by which an orthogonal basis is constructed from data and is frequently used to perform order reduction. In the past, it has been used in a wide range of machine learning applications, from image de-noising to gene processing \cite{Zhang_Dong_Zhang_Shi_2010} \cite{Yeung_Ruzzo_2001}. From a data matrix $\mathbf{U}$ with $n$ columns, each consisting of $\mathbf{d}$ at different points in time, the covariance matrix $\mathbf{C}$ is calculated as:
\begin{equation}
    \mathbf{C} = \frac{1}{n-1}\mathbf{U}^T\mathbf{U}.
\end{equation}
Each POD base $\mathbf{\phi}_i$ is an eigenvector of the covariance matrix. The columns of the POD matrix $\mathbf{\Phi}$ are the first $D_0 = P$ POD bases. A linear combination of POD bases produces an approximation of the data vector $\mathbf{d}$. Assigning $\mathbf{p}$ to the vector denoting the combination of POD bases, it's possible to write:
\begin{equation} \label{eq:PODforward}
    \mathbf{d} = \mathbf{\Phi}\mathbf{p}.
\end{equation}
Therefore, a calculation of $\mathbf{p}$ allows a calculation of $\mathbf{d}.$

POD produces an ordered list of most significant axes, meaning the first POD base is the data axis with the highest variance, and the coordinate along this axis is the single most important number for distinguishing different states from each other. POD is therefore useful for creating an orthogonal basis that is able to distinguish points in data space with as few inputs as possible. This quality makes it a good candidate for understanding which electrode placements measure the data as effectively as possible, as well as distinguishing points in data space as with as few measurements as possible.

\section{Methods}
\subsection{Electrode Placement}
The electrode placement algorithm in this work draws heavily from the sensitivity volume method, an algorithm which calculates which electrodes should be used to make the $D_0$ measurements with the highest signal-to-noise ratio \cite{onsager2024sensitivity}. The main concept behind the sensitivity volume method is that a high quality set of measurements is highly linearly independent, which is to say that each measurement is adding as much non-redundant information as possible. In the original sensitivity volume method, the sensitivity $S$ measures the linear independence of the rows of the Jacobian, therefore quantifying how well a measurement set is able to span each location in the finite element mesh. The sensitivity is calculated as:
\begin{equation}
    S = \sqrt{\det(\mathbf{J}\mathbf{J}^T)}.
\end{equation}

The goal of an electrode placement algorithm using the POD is to find the set of measurements that most accurately span the POD bases, or $\mathbf{p}$. Spanning $\mathbf{p}$ is superior to spanning $\mathbf{m}$ because the POD is based on data that has already been collected, indicating which parts of the ROI are most important to measure when distinguishing points in dataspace. 

\begin{figure}[h]

\begin{subfigure}{0.33\textwidth}
\includegraphics[width=0.9\linewidth]{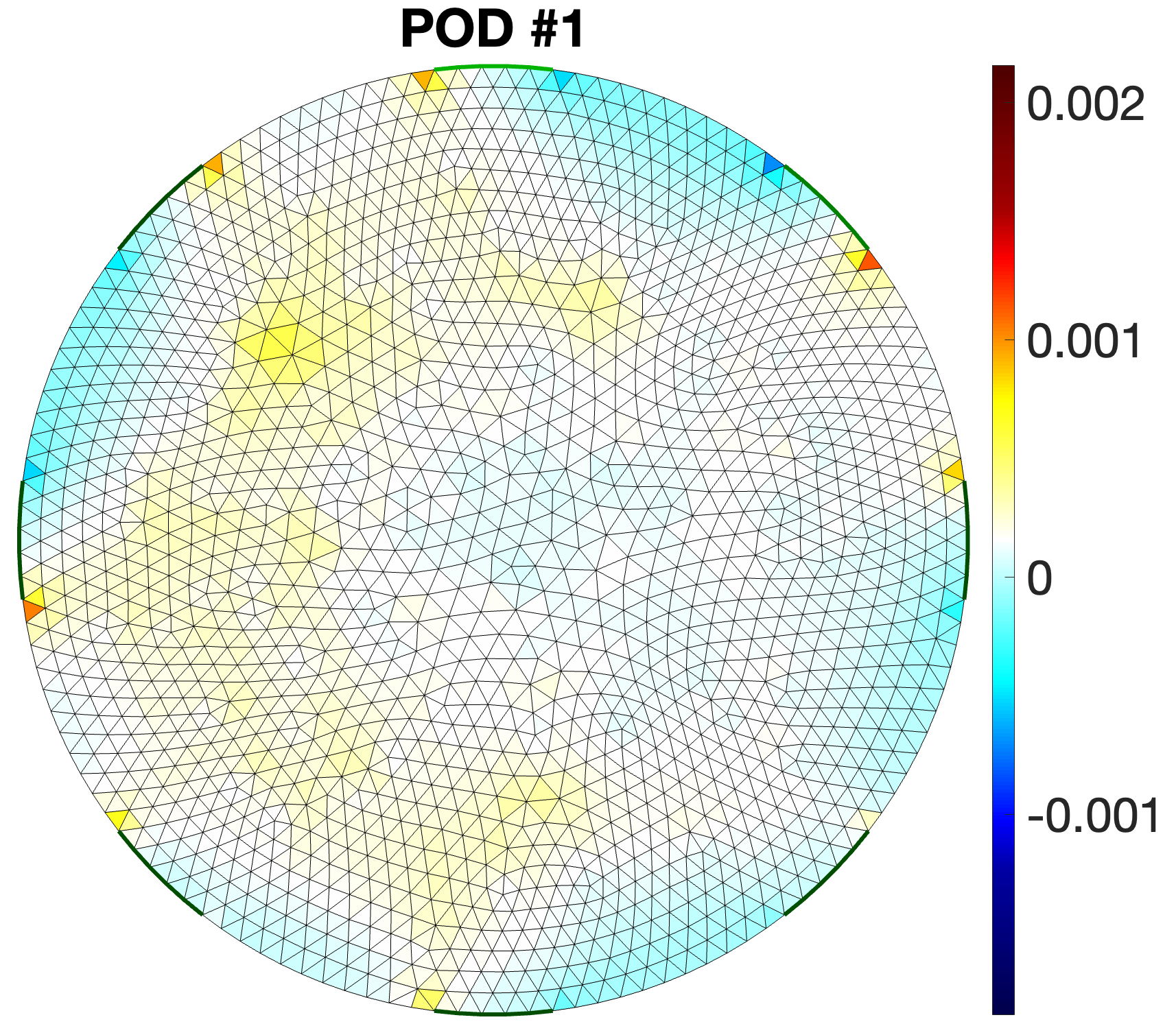} 
\caption{Mesh $\phi_1$}
\label{fig:subim1}
\end{subfigure}
\begin{subfigure}{0.33\textwidth}
\includegraphics[width=0.9\linewidth]{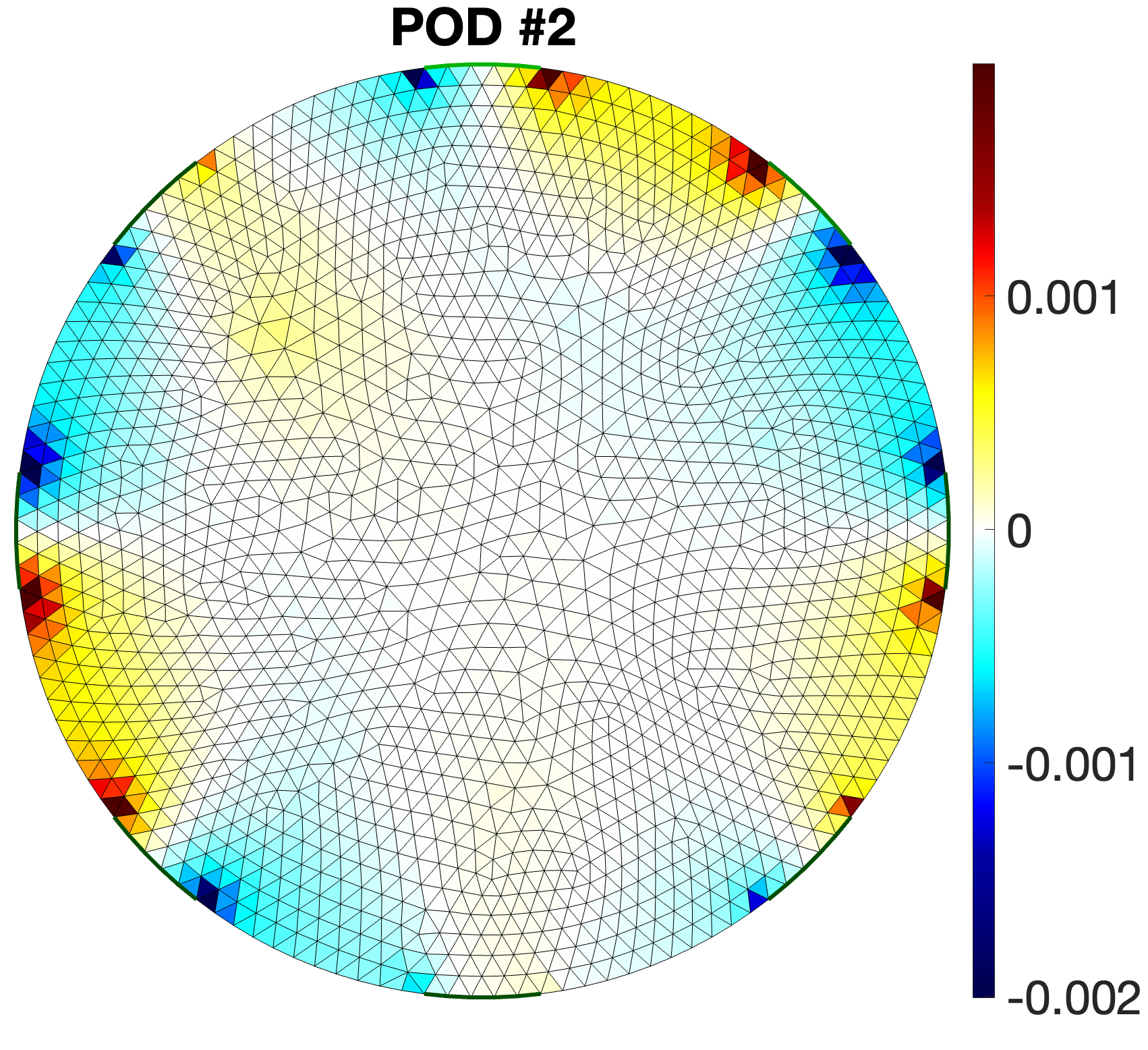} 
\caption{Mesh $\phi_2$}
\label{fig:subim1}
\end{subfigure}
\begin{subfigure}{0.33\textwidth}
\includegraphics[width=0.9\linewidth]{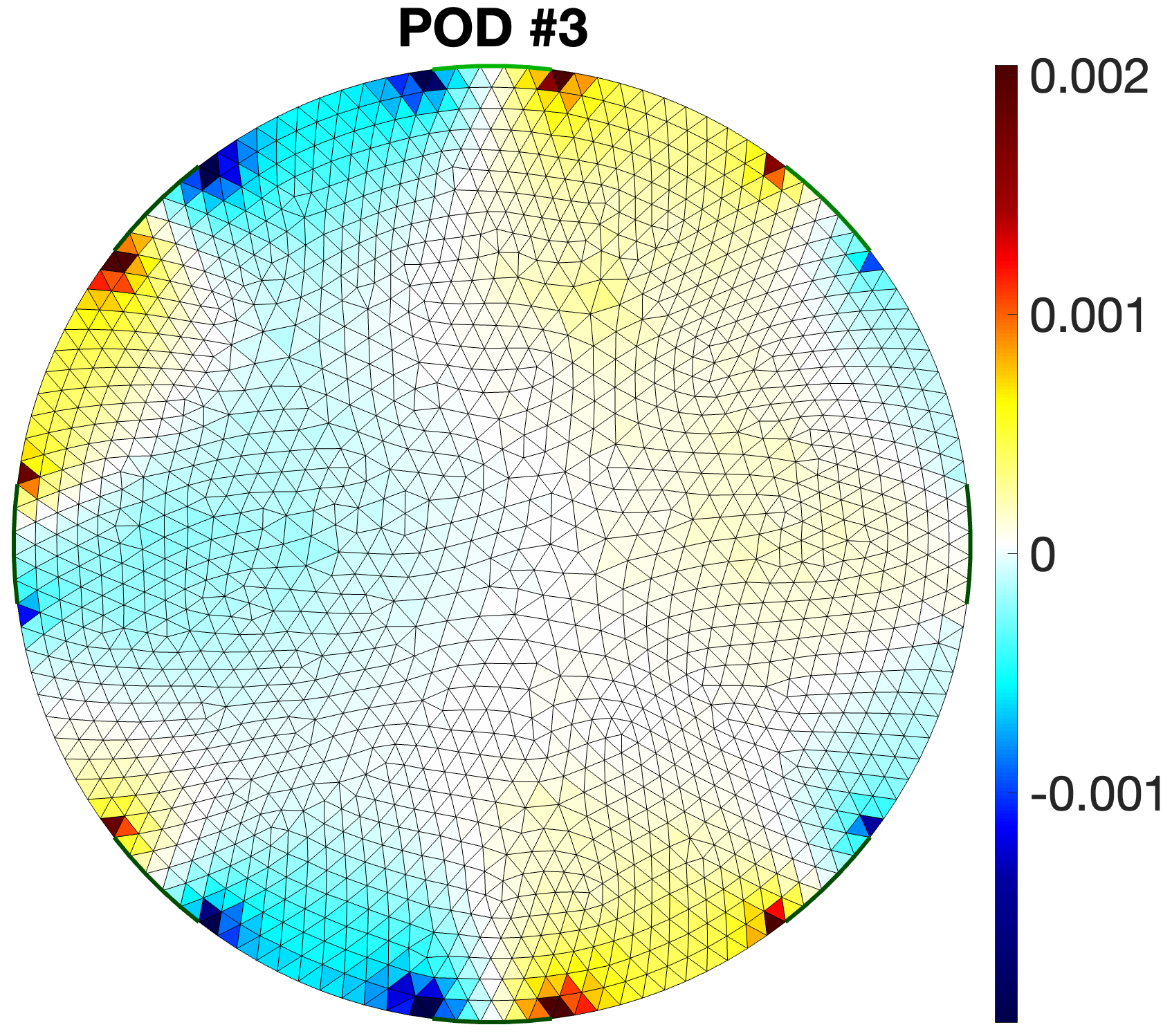} 
\caption{Mesh $\phi_3$}
\label{fig:subim1}
\end{subfigure}

\caption{The mesh projections of the first three POD bases. Each image has the same color bar range; note the first basis is smaller in magnitude. Each circle is a cross-section of the forearm, with the left of the circle lining up roughly with the ulna. Interestingly, the edges of the projections have a large magnitude, indicating that much of the difference in data comes from electrode contact. }
\label{fig:PODMesh}
\end{figure}

When optimizing electrode placement, it is important to be able to consider measurements that were not in the original dataset. For this reason, it's necessary to project the POD bases onto the finite element mesh in order to measure how well arbitrary measurements span the $\mathbf{p}.$ This projection takes the form of a mesh POD matrix $\mathbf{\Phi}_M,$ in which each column describes which areas of the ROI are measured by the corresponding POD basis. It is calculated with:
\begin{equation}
    \mathbf{\Phi}_M = \mathbf{J}^T \mathbf{\Phi}
\end{equation}
It's possible to visually inspect the POD bases by plotting the columns of $\mathbf{\Phi}_M$, as done in Fig. \ref{fig:PODMesh}.

\begin{figure}
    \centering
    \includegraphics[width=0.5\linewidth]{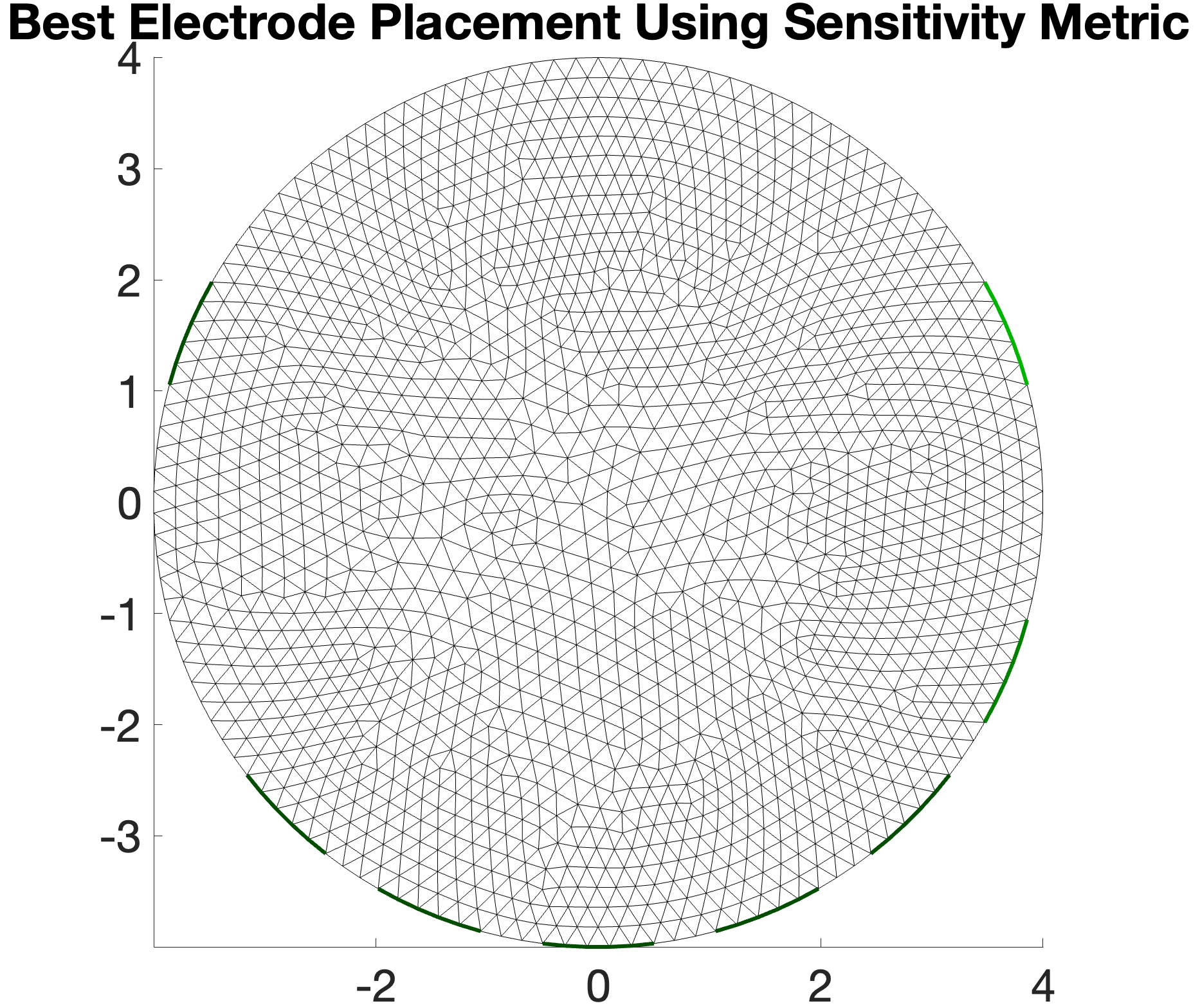}
    \caption{An image of the best electrode configuration according to the electrode placement method used in this paper. Each of the dark lines on the sides is an electrode.}
    \label{fig:bestelec}
\end{figure}

Once each POD basis is projected into the mesh space, the electrode optimization algorithm is performed. This electrode optimization algorithm works by placing 16 potential electrode locations evenly spaced around the outside of the circle, and then for each combination of eight electrodes, the POD sensitivity $S_\phi$ is calculated assuming the "skip" measurement protocol:
\begin{equation}
    S_\phi = \sqrt{\det(\mathbf{J}\mathbf{J}^T \mathbf{\Phi})}
\end{equation}
The set of electrodes with the highest POD sensitivity was selected as the optimal arrangement, and is shown in Fig. \ref{fig:bestelec}
Data was taken using both this configuration and with the electrodes evenly spaced, and the results are discussed in section \ref{s:evaluation}.
\subsection{Data Projection}
The other novel part of this work regards a data projection algorithm, which allows for the compensation of electrodes not taking accurate data. The goal is to take a subset of measurements all using electrodes taking accurate data $\mathbf{d}'$ and project it to $\mathbf{d}''$ with the rank of $\mathbf{d}$ such that models trained on $\mathbf{d}$ can still function accurately. This is accomplished by solving for $\mathbf{p}'$ given $\mathbf{d}'$ then finding $\mathbf{d}''$ using (\ref{eq:PODforward}). The first step is done by calculating:
\begin{equation} \label{eq:d'}
    \mathbf{p}' = \mathbf{\Phi}'{^-1} \mathbf{d'}
\end{equation}
where $\mathbf{\Phi}'$ is a truncated form of $\mathbf{\Phi},$ only containing the first $D' = |\mathbf{d'}|$ POD bases, and only the rows of those that correspond to the measurements in $\mathbf{d}'.$ This is so that the matrix is square, making it invertible, and also ensuring that $\mathbf{p}'$ is unique. The next goal is to turn $\mathbf{p}'$ into a data vector with the same format as the vectors that trained the machine learning model currently being used. This is done with:
\begin{equation} \label{eq:d''}
    \mathbf{d}'' = \mathbf{\Phi}''\mathbf{p}'
\end{equation}
were $\mathbf{\Phi}''$ contains each row, but is still truncated to the first $D'$ POD bases, meaning $|\mathbf{d}''| = |\mathbf{d}|.$ Combining (\ref{eq:d'}) and (\ref{eq:d''}):
\begin{equation}
    \mathbf{d}'' = \Phi'' (\Phi'^{-1} \mathbf{d}')
\end{equation}This method essentially projects data with fewer electrodes into data with more electrodes, allowing for real-time compensation for disconnected or poor-signal electrodes.

\section{Evaluation} \label{s:evaluation}
\subsection{Electrode Placement}
\begin{figure}[h!]

\begin{subfigure}{0.5\textwidth}
\includegraphics[width=0.9\linewidth]{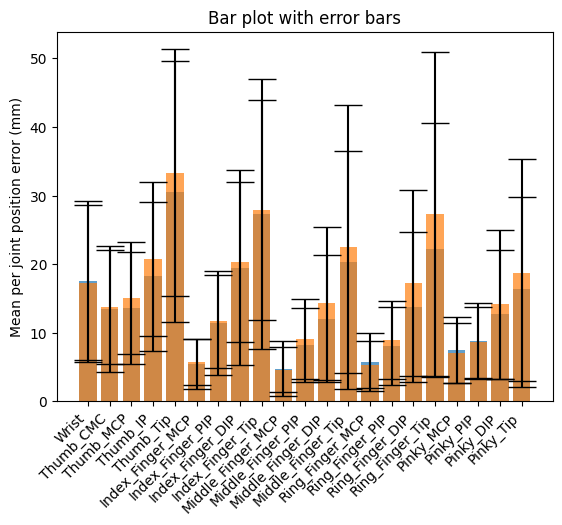} 
\caption{Evenly Spaced}
\label{fig:subim1}
\end{subfigure}
\begin{subfigure}{0.5\textwidth}
\includegraphics[width=0.9\linewidth]{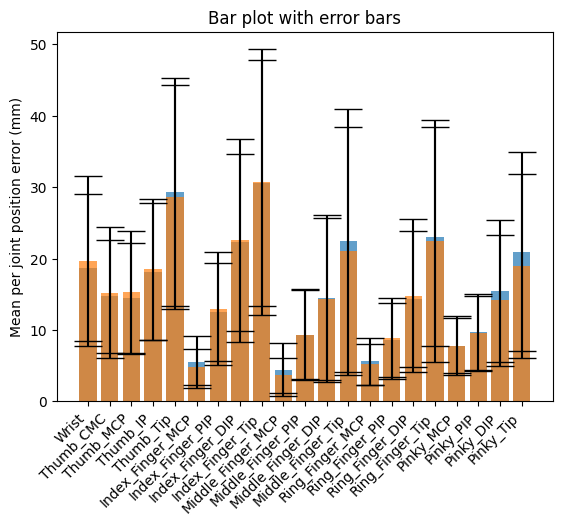} 
\caption{Algorithm-Determined Placement}
\label{fig:subim1}
\end{subfigure}

\caption{The MPJPE for each possible different hand poses for evenly spaced electrodes (a) and algorithm-determined electrode placement (b), showing the error for each pose. Blue is trained on 20 minute session and tested on 10 minute session, while orange is the opposite. The two graphs each look quite similar, meaning the two electrode placements have little difference in error.}
\label{fig:PlacementError}
\end{figure}

It is necessary to evaluate the performance of both the electrode placement and data projection algorithms. This was done using the mean per-joint position error (MPJPE) evaluation pipeline of the original EITPose paper \cite{Kyu_Mao_Zhu_Goel_Ahuja_2024}. Two sessions of data were taken with each electrode configuration, one of roughly 20 minutes, the other of roughly ten. These two conditions are compared in Fig. \ref{fig:PlacementError}
The error charts look very similar to each other, meaning that electrode placement seems to have little effect on the accuracy of EITPose. The MPJPE for the evenly spaced electrodes when trained on 20 minutes and tested on 10 is 14.1 mm, while for the algorithm-determined electrode placement it is 15.3 mm. Since this is an overall low difference in error and the charts in Fig. \ref{fig:PlacementError}, look similar, it appears that electrode placement matters little. Based on this fact, it seems that most of the information in the EITPose data is contained in how well the electrodes are able to make skin contact at different hand positions.

\subsection{Data Projection}
\begin{figure}
    \centering
    \includegraphics[width=0.5\linewidth]{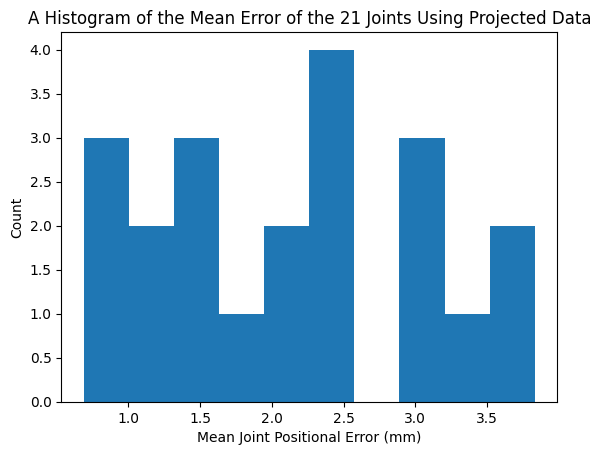}
    \caption{A histogram of each joint's error when tested on projected data. The distribution is relatively uniform, but the data projection algorithm is able to communicate some joints better than others.}
    \label{fig:hist}
\end{figure}

It's also necessary to calculate the error for the data projection method. In this case, a post-processing algorithm was run on previously collected data from the EITPose paper that projected each frame of data as if the first electrode was invalid. The model was then trained on the set of data before projection, and tested after projection. This yielded a MPJPE of 2 mm, and a histogram of each of the joints' errors can be seen in Fig. \ref{fig:hist}.
With future evaluation, it would be potentially useful to collect much more data of the two different electrode placements to see if there is any difference between the placements. It would also be interesting to try a few more electrode placements -- perhaps placement does indeed matter very little, or maybe the two placements in this work happen to be equally good. It would also be interesting to do an ablation study by projecting out each of the eight different electrodes to see which ones have the highest error.
\\
\\
\section{Conclusion}
This work presents a new method of electrode placement alongside a new method for data projection. The electrode placement algorithm proved to have little impact on hand position error, but the data projection algorithm is able to account for low-signal electrodes with very little error introduced. This is promising for any application that uses EIT for live applications and has a large store of data on which to perform POD. This includes any future applications for EITPose, especially in applications like watch bands where contacts may make poor contact at any given time. It is also relevant for applications of EIT with large trained models where electrode reduction is desired. For example, this algorithm could project data from 6 electrodes up to 8 electrodes, allowing complex models to be trained with large numbers of electrodes, with consumer devices with fewer electrodes projecting up their data. This paper ultimately represents the introduction of a highly useful data projection algorithm that can be used to improve the performance of future EIT devices.

\bibliographystyle{unsrt}  
\bibliography{references}

\end{document}